\documentclass{mem}
\usepackage{natbib}
\usepackage{txfonts}
\usepackage{balance}
\usepackage{graphicx}
\usepackage[a4paper]{hyperref}
\idline{75}{282}

\newcommand{\chem}[2]{$\mathrm{^{#1}#2}$}

\begin{document}
\title{Evolution of the main observables in core-collapse supernova events: a radiation-hydrodynamical modelling}

\author{
M.L. \,Pumo\inst{1,2} 
\and L. \,Zampieri\inst{1}}

\offprints{M.L. Pumo}

\institute{
INAF - Osservatorio Astronomico di Padova, Vicolo dell'Osservatorio 5, I-35122
Padova, Italy
\and
INAF - Osservatorio Astrofisico di Catania, Via S. Sofia 78,
I-95123 Catania, Italy
}
\authorrunning{Pumo \& Zampieri}

\titlerunning{Radiation-hydrodynamical modelling of CC-SNe}
   \subtitle{}

\abstract{
With the aim of clarifying the nature of the supernova events from stars having initial (at the main sequence) masses larger than $\sim 8$-$10\,M_{\odot}$, we have developed a specifically tailored radiation hydrodynamics Lagrangian code, that enables us to simulate the evolution of the main observables (light curve, evolution of photospheric velocity and temperature) in these events. The code features and some test-case simulations as well as the possible applications of the code are briefly discussed.

}

\maketitle


\section{Introduction}

It is widely accepted that core-collapse supernovae (CC-SNe) represent the final explosive evolutionary phase of stars having initial (i.e. at main sequence) masses larger than $\sim 8$\,-\,$10\,M_{\odot}$ \citep[see e.g.][]{p5,p7,p14}. In addition to their intrinsic interest, these explosive events are relevant to many astrophysical issues associated, for example, with the nucleosynthesis processes of intermediate and trans-iron elements, the physical and chemical evolution of the galaxies, the production of neutrinos, cosmic rays and gravitational waves \citep[see e.g.][]{p1,p3,p6,p8,p12}. Moreover, CC-SNe seem to be particularly promising to measure cosmological distances, in addition to the well known method using type Ia SNe \citep[see e.g.][and references therein]{p17bis}.

\begin{figure*}[t!]
\resizebox{\hsize}{!}{\includegraphics[clip=true]{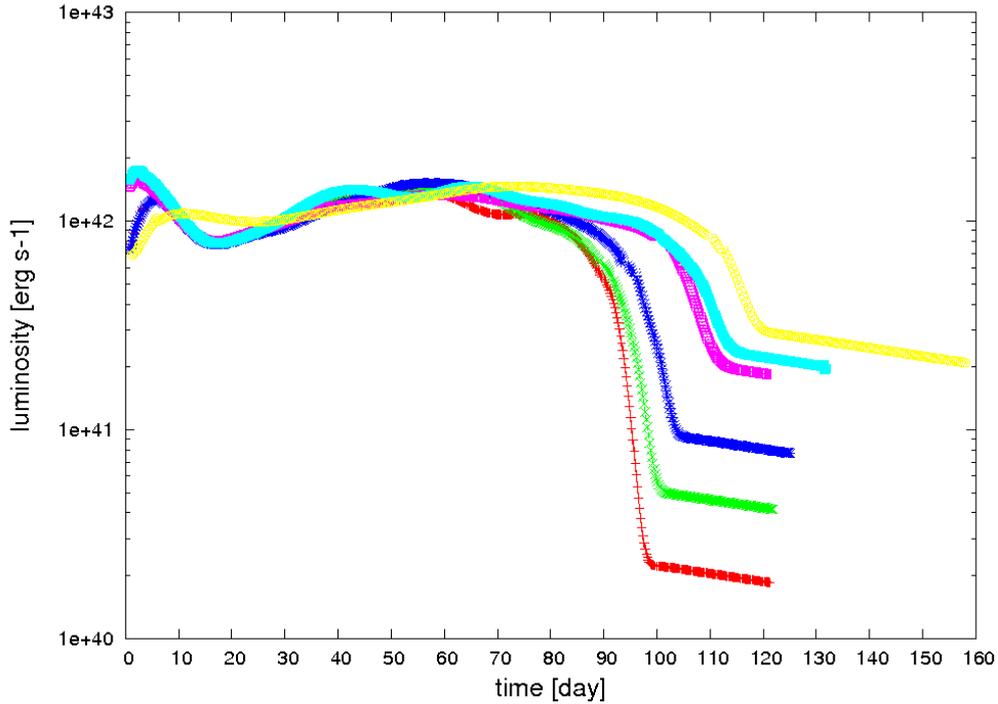}}
\caption{\footnotesize Effects of varying the \chem{56}{Ni} mass on the light curve for models having the same total mass ($16\,M_{\odot}$), initial radius ($3\times10^{13}$cm) and total (kinetic plus thermal) energy ($1$ FOE $\equiv$10$^{51}$\,ergs). The tracks correspond to \chem{56}{Ni} masses equal to $0.005\,M_{\odot}$ (red line), $0.010\,M_{\odot}$ (green line), $0.020\,M_{\odot}$ (blue line), $0.045\,M_{\odot}$ (purple line), $0.055\,M_{\odot}$ (cyan line), and $0.070\,M_{\odot}$ (yellow line).}  
\label{fig_2}
\end{figure*}
\begin{figure*}[t!]
\resizebox{\hsize}{!}{\includegraphics[clip=true]{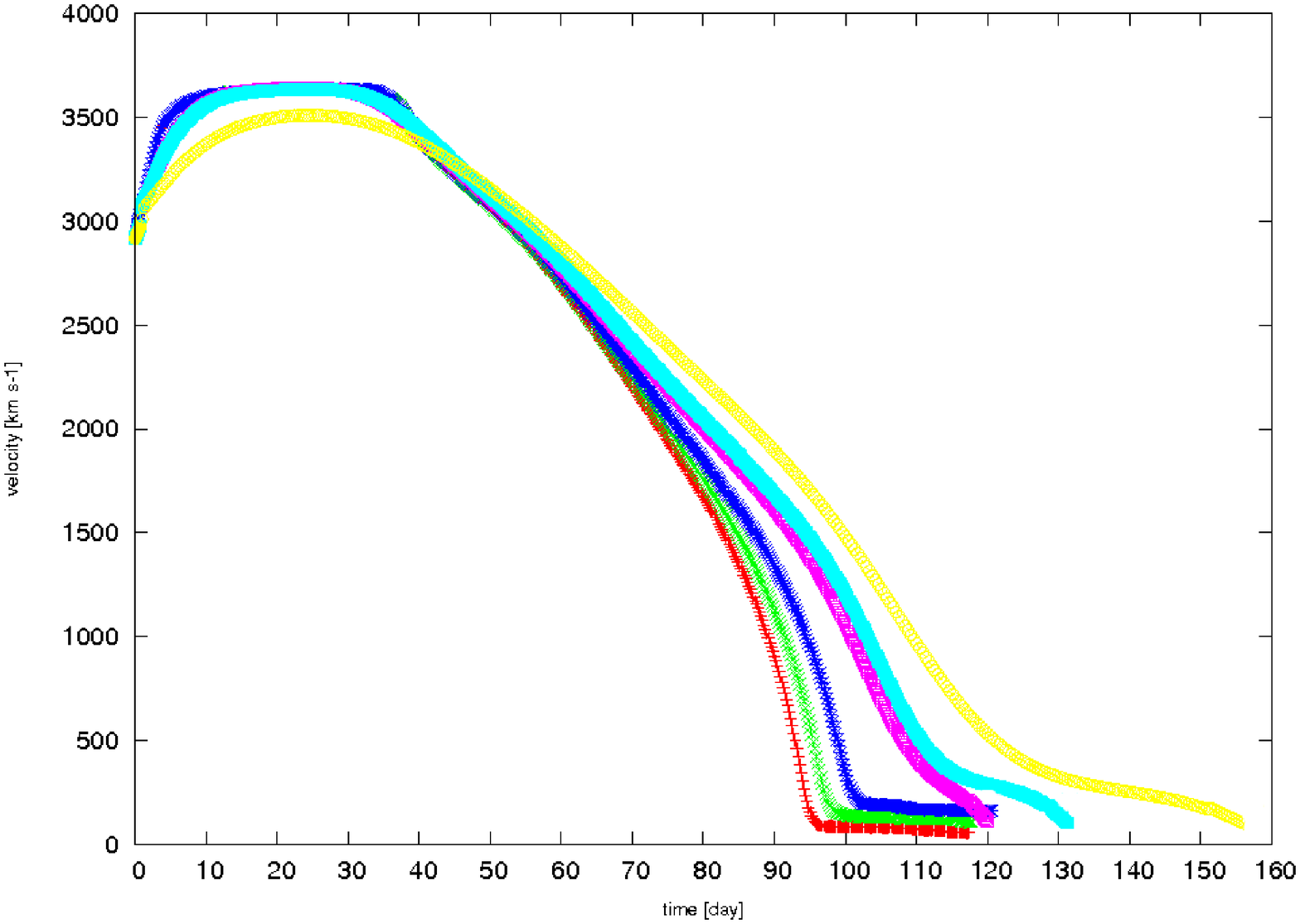}}
\caption{\footnotesize Same as Figure \ref{fig_2}, but for the evolution of the photospheric velocity.}  
\label{fig_3}
\end{figure*}

Despite the importance of these explosive events in astrophysics, there are still ambiguities and uncertainties, linked to the extreme variety of CC-SNe displays (various energetics and different amounts of ejected material, leading to heterogeneous light curves, spectra and evolution of photospheric velocity and temperature) and related to the uncertainties in the modelling of stellar evolution and explosion mechanism \citep[see e.g.][and references therein]{p4,p9,p13}. In particular the exact link between the physical properties of the SN explosion (ejected mass, explosion energy, stellar structure and composition at the explosion) and the observational features is far from being well established, and a ``self-consistent'' description of CC-SN events (from the quiescent evolutionary phases up the post-explosion evolution) is still missing.

In this context, the creation of specific modelling tools that link progenitor evolution and explosion models to the main observables (i.e. light curve, evolution of photospheric velocity and temperature) of CC-SNe is of primary importance for improving our knowledge about the nature of the CC-SN events, and the development of the new, radiation-hydrodynamics code, described in the next section, represents a key step in this direction.


\section{Code description and simulations}

The code, which solves the equations of relativistic, radiation hydrodynamics for a self-gravitating matter fluid interacting with radiation, is a new and improved version of the one reported in \citet[][]{p15}, \citet[][]{p16} and \citet[][]{p2}. 

Major strengths of the new version of the code are (see also \citealt{p10} and Pumo \& Zampieri 2010, in prep): 
\begin{itemize}
 \item[-] an accurate treatment of radiative transfer in all regimes (from completely optically thick to optically thin);
 \item[-] the coupling of the radiation moment equations with the equations of relativistic hydrodynamics, adopting a fully implicit Lagrangian finite difference scheme, during all the post-explosive phases;
 \item[-] the computation of the evolution of the ejecta and the emitted luminosity taking into account the gravitational effects due to the compact remnant, enabling the code to deal with the fallback of material onto the compact remnant and, consequently, avoiding to overestimate the ejection of \chem{56}{Ni}.
\end{itemize}

These distinctive features make it possible to follow the post-explosion evolution of a CC-SN event in its ``entirety'', from the breakout of the shock wave at the stellar surface up to the so-called nebular stage, in which the ejected envelope has recombined and the energy budget is dominated by the radioactive decays of the heavy elements synthesized in the explosion. 

Consequently, the code is able to simulate the post-explosion evolution of the main observables in CC-SN events for a given set of initial parameters (kinetic and thermal energy, envelope mass, radius, amount and distribution of \chem{56}{Ni}) that are linked to the physical conditions of the CC-SN progenitor at explosion.


\section{Future developments}

We are starting to compute fine grids of models evolved from post-explosion configurations and to carry out detailed investigations of the evolution of CC-SNe, in order to determine how the physical conditions of the CC-SN progenitor at the explosion affect the evolution of the main observables (see Fig.s \ref{fig_2} and \ref{fig_3} for the preliminary results of some of these calculations). 

Our long term goal is the development of a sort of ``CC-SNe Laboratory'' interfacing our new relativistic, radiation-hydrodynamics code with stellar evolution, nucleosynthesis and spectral synthesis codes. This will allow us to describe the evolution of a CC-SN event in a ``self-consistent'' way from the evolutionary stages preceding the main sequence up to the post-explosive phases as a function of initial mass, metallicity and mass loss history of the CC-SN progenitor.
 

\begin{acknowledgements}
M.L.P. acknowledges the financial support by the Bonino-Pulejo Foundation. The TriGrid VL project and the INAF - Osservatorio Astronomico di Padova are also acknowledged for computer facilities.
\end{acknowledgements}

\bibliographystyle{aa}

\end{document}